\documentclass[copyright,creativecommons]{eptcs}
\providecommand{\event}{Gandalf 2016} 

\usepackage{alltt}
\usepackage{graphicx}
\usepackage{color}
\usepackage{nicefrac}
\usepackage{subfigure}
\usepackage{amsmath}
\usepackage{amssymb}
\usepackage{float}
\usepackage{framed}
\usepackage{stmaryrd} 
\usepackage{tikz}
\usepackage{tkz-graph}
\usepackage{lmodern}
\usepackage{fancyvrb, relsize}
\usepackage{listings}
\usepackage{subfigure}
\usepackage{float}
\usepackage{multirow}
\usepackage{paralist}
\usepackage{alltt}
\usepackage{array}
\usepackage{xspace}

\title{Partial Solvers for Parity Games:\\ Effective Polynomial-Time Composition}
\author{Patrick Ah-Fat and Michael Huth
\institute{Imperial College London\\ London, UK}
\institute{Department of Computing}\\
\email{\{patrick-ah-fat14, m.huth\}@imperial.ac.uk}
}
\def\titlerunning{Partial Solvers for Parity Games}
\def\authorrunning{P.\ Ah-Fat \& M.\ Huth}
\begin{document}
\maketitle

\newcommand{\qed}{{\bf \ QED.}}

\newtheorem{theorem}{Theorem}
\newtheorem{definition}{Definition}
\newtheorem{lemma}{Lemma}
\newtheorem{example}{Example}

\marginparwidth=3cm
\newcommand{\rank}[1]{{\mathsf r}({#1})}
\newcommand{\while}[1]{{\mathsf {while}}({#1})}
\newcommand{\chain}[1]{{\while{#1}}}
\newcommand{\hide}[1]{}
\newcommand{\Inf}[1]{\mathsf{Inf}({#1})}
\newcommand{\Col}[1]{\mathsf{C}({#1})}
\newcommand{\Nat}{\mathbb{N}}
\newcommand{\size}[1]{{\mid \! {#1}\! \mid}}
\newcommand{\psolb}{{\tt psolB}\xspace}
\newcommand{\psolc}{{\tt psolC}\xspace}
\newcommand{\attr}[3][p]{{{\mathsf{Attr}}_{{#1}}[{#2},{#3}]}}
\newcommand{\win}[2]{{{\mathsf{Win}}_{{#1}}[{#2}]}}
\newcommand{\lift}[1]{\mathsf{lift}({#1})}
\newcommand{\lifted}[1]{\mathsf{lifted}({#1})}
\newcommand{\call}[1]{\mathsf{call}({#1})}
\newcommand{\scc}{{scc}}
\newcommand{\pp}{{pp}}
\newcommand{\fa}{{f\!a}}
\newcommand{\gfa}{{g\!f\!a}}
\newcommand{\ari}{{ari}}
\newcommand{\mss}{{m_{ss}}}
\newcommand{\mscc}{{m_{scc}}}
\newcommand{\erfa}{{er_{{}\!\fa}}}
\newcommand{\ersd}{{er_{{}\!sd}}}
\newcommand{\psolone}{{ps_1}}
\newcommand{\psoltwo}{{ps_2}}
\newcommand{\psolthree}{{ps_3}}
\newcommand{\psolfour}{{ps_4}}
\newcommand{\psolfive}{{ps_5}}
\newcommand{\gamesizeone}{0.45}
\newcommand{\gamesizetwo}{0.45}
\newcommand{\gamesizethree}{0.3376}
\newcommand{\gamesizefour}{0.45}
\newcommand{\nav}{\!\bullet\!}
\newcommand{\preforder}[1]{\,{\preceq_{#1}}\,}
\newcommand{\mybibitem}{\vspace{-0.325mm}\bibitem}

\graphicspath{{./figures/}}
 
\begin{abstract}
Partial methods play an important role in formal methods and
beyond. Recently such methods were developed for parity games, where
polynomial-time partial solvers decide the winners of a subset of
nodes. We investigate here how effective polynomial-time partial
solvers can be by studying interactions
of partial solvers based on generic composition
patterns that preserve polynomial-time
computability. We show that use of such composition patterns
discovers new partial solvers~--~including those that merge
node sets that have the same but unknown winner~--~by studying games
that composed partial solvers can neither solve nor simplify. We
experimentally validate that this data-driven approach to refinement
leads to polynomial-time partial solvers that can solve \emph{all}
standard benchmarks of structured games. For one of these
polynomial-time partial solvers not even a sole random game from a few
billion random games of varying configuration was found
that it won't solve completely.
\end{abstract}

\section{Introduction}
\label{section:introduction}
Parity games are two-player games on directed graphs that are
determined \cite{Most91,EJ91,Zie98}. Parity games have several
applications, including as back-ends in formal methods.
The exact computational complexity for finite parity games has been an open problem for over 20 years: deciding which player wins a node in a
parity game is in UP$\cap$coUP \cite{Jurdzinski98} and the fastest known
algorithms run in sub-exponential time in the size of games, see e.g.\ \cite{JPZ06,DBLP:conf/csl/Schewe08}. 
Some types of parity games have polynomial-time solutions. Bounding the index of games~--~i.e.\ the largest color of a
game~--~by a fixed number, Zielonka's algorithm based on the whole-set rule \cite{Friedmann11}
becomes polynomial time. Or 
we may bound a descriptive complexity measure: parity games with bounded DAG-width \cite{BDHK06}, tree-width \cite{BDHK06,DBLP:journals/corr/abs-1112-0221} or entanglement \cite{DBLP:journals/tcs/BerwangerGKR12}
can be solved in polynomial time.

Algorithms that solve parity games do so using specific
mechanisms, for example strategy improvement \cite{VJ00} or progress
measures \cite{Jur00}. But it seems not feasible to let such mechanisms
interact in iterative computations, even though this might speed up
solving time. The difficulty is that such mechanisms operate over very different
views of games and their complexity;
for example, how might one use a strategy-improvement step (which updates one player's strategy) to increase
a progress measure (an element in a specific complete lattice)?

Partial solvers \cite{DBLP:conf/fossacs/HuthKP13,DBLP:journals/corr/HuthKP14} have been proposed as algorithms that can
solve parts of a parity game but not necessarily all of such a
game. Such algorithms are designed to run in polynomial time, and this is relatively easy to obtain. The harder part is to understand which parity games are solved completely by a given partial solver. Partial solvers are related to known
static analyses such as priority propagation
(see e.g.\ \cite{Friedmann10}), that may decrease colours of nodes. Extant work has shown
the feasibility of using partial solvers \cite{DBLP:conf/fossacs/HuthKP13,DBLP:journals/corr/HuthKP14}, yet they don't
completely solve some benchmarks of structured games and they don't
solve many randomly generated games. In this paper, we ask whether
partial solvers can improve their
effectiveness through interaction.

Addressing this question seems feasible as all these methods share a common
view of the complexity of a finite game~--~say the number of nodes
plus the number of edges plus the sum of all colours of all
nodes. This common view allows us to think of static analyses, let us mention color reductions based on abstract Rabin index computations \cite{HKP15}, as
partial solvers as well and to then \emph{compose} partial solvers to
improve their effectiveness.
This discussion leads us to consider whether there are
simple, generic, yet effective composition patterns for partial 
solvers that preserve polynomial-time computability, allow us to
manually discover new partial solvers, and that
can solve \emph{all} standard benchmarks of structured games
and only very rarely do not completely solve a
  randomly generated game. The main contribution of this paper is
to  provide such an approach and experimental evidence that such aims are realizable.

\paragraph{{\bf Outline of paper:}} We review background in
Section~\ref{section:background}, develop our composition approach
for partial solvers in Section~\ref{section:composition}, and show how
its use leads to data-driven refinement of  partial solvers in
Section~\ref{section:refinement}. In
Section~\ref{section:experiments}, we report our experimental and
validation work for our approach and its newly discovered
partial solvers. Related work is discussed in
Section~\ref{section:related}, further insights are discussed in
Section~\ref{section:discussion}, and Section~\ref{section:conclusion}
concludes the paper. We refer to \cite{TR} for proofs, a formal presentation of our approach, and further
experimental details not provided in this paper.

\section{Background}
\label{section:background}
We define key concepts of parity games, review some partial solvers and static analyses for such games, and fix technical notation used in this paper.
We write $\Nat$ for the set $\{0,1,\dots\}$ of natural numbers. 
A parity game $G$ is a tuple $(V,V_0,V_1,E,c)$, where $V$ is a
set of nodes partitioned into possibly empty node sets $V_0$
and $V_1$, with an edge relation $E\subseteq V\times V$ that contains no dead-ends (i.e.\ for all
$v$ in $V$ there is a $w$ in $V$ with $(v,w)$ in $E$), and a colouring
function $c\colon V\to \Nat$. Throughout, we write $p$ for one of $0$ or $1$ and $1-p$ for the other
player.  Nodes in $V_0$ are owned by player $0$, nodes in $V_1$ are
owned by player $1$. We write $owner(v)$ to denote the $p$ for which
$v$ is in $V_p$. In figures, $c(v)$ is written within
nodes $v$, nodes in $V_0$ are depicted
as circles and nodes in $V_1$ as squares. 
For a relation $\rho\subseteq A\times B$ and $X\subseteq A$
we write $X\nav\rho$ for set $\{ b\in B\mid \exists a\in X\colon  (a,b) \in \rho\}$, whereas
$\rho\nav Y$ denotes set $\{a\in A\mid \exists b\in Y\colon (a,b)\in \rho\}$ for $Y\subseteq
B$; we will abuse this notation for singleton $X$ and $Y$ as in $v\nav
E$
or $E\nav v$ in a parity game.
Below we write $\Col G$ for the set of colours in
game $G$, i.e.\
\[
\Col G = \{c(v)\mid v\in V\}
\]

\noindent We write $x\% 2$ for $x$ modulo
$2$ for an integer $x$. For each $p$ in $\{0,1\}$, the preference ordering $\preforder p$ on $\Col G$ is given by $c_1\preforder p c_2$ iff ($c_1\% 2 = p$ and $c_2\%2 = 1-p$) or ($c_1$ and $c_2$ have parity $p$ and $c_1\leq c_2$) or ($c_1$ and $c_2$ have parity $1-p$ and $c_2\leq c_1$).

A play from some node $v_0$ results in an infinite play $\pi =
v_0v_1\dots$ in $(V,E)$ where the player who owns $v_i$ chooses the
successor $v_{i+1}$ such that $(v_i,v_{i+1})$ is in $E$.  
Let $\Inf \pi$ be the set of colours that occur in $\pi$ infinitely
often:
\[
\Inf \pi = \{ k\in\Nat \mid \forall j\in \Nat \colon \exists
i\in \Nat \colon i>j \mbox{ and } k=c(v_i) \}
\]

\noindent Player $0$ wins play $\pi$ iff $\min \Inf \pi$ is even; otherwise
player $1$ wins play $\pi$.
A strategy for player $p$ is a total function $\sigma_p\colon V^*\cdot V_p\to V$
where the pair $(v,\sigma_p(w\cdot v))$ is in $E$ for all $v$ in $V_p$ and $w$ in $V^*$.
A play $\pi$ conforms with $\sigma_p$ if for every finite prefix $v_0\dots v_i$ of $\pi$ 
with $v_i$ in $V_p$ we have $v_{i+1}=\sigma_p(v_0\dots v_i)$.  

Parity games are determined \cite{Most91,EJ91,Zie98}:
(i) node set $V$ is the disjoint union of two, possibly empty,
sets $\win 0G$ and $\win 1G$, the
winning regions of players $0$ and $1$ (respectively) in $G$; and (ii)
there are (memoryless) strategies $\sigma_0$ and $\sigma_1$ such that
all plays beginning in $\win 0G$ and conforming with $\sigma_0$ are won by 
  player $0$, and all plays beginning in $\win 1G$ and conforming with $\sigma_1$ are won by
  player $1$.
By abuse of language,
$\emptyset$ is also a parity game with no nodes.  

We define the \emph{rank} of parity game $G$ as
$$
\rank G = \size V + \size E + \sum_{v\in V} c(v)
$$

\noindent We write $\attr[p]{G}{X}$ for the attractor
of node set $X$ for player $p$, which computes the 
alternating reachability of $X$ for that player in the game graph of
$G$ (see e.g.\ Definition~1 in \cite{DBLP:conf/fossacs/HuthKP13}). It
is well known that $\attr[p]{G}{X}$ is contained in the winning region
$\win pG$ whenever $X\subseteq \win pG$.
The color of a finite path or cycle $P$ in the directed graph $(V,E)$ is defined to be $\min \{
c(v)\mid v\hbox{ is on }P\}$. A subset $C\subseteq V$ of a directed
graph is called a (maximal) strongly connected component, denoted by
SCC, if for all $v,w$ in $C$ there is a path in $(V,E)$ from $v$ to
$w$; and if there is no strict superset of $C$ in $(V,E)$ with that property.
\begin{example}
\label{example:pg}
For parity game $G$ on the right in Figure~\ref{fig:mss}, we have $\win 1G = \emptyset$ and $\win 0G = V$.
The (memoryless) strategy $\sigma_0$ with $\sigma_0(v_4) = v_{19}$
is a winning strategy for player $0$ on $\win 0G$. 
\end{example}

We present partial solvers and static analyses for parity games, some of them already in a form suitable for the composition patterns developed in this paper.
All these partial solvers and static analyses preserve the winning regions of the (remaining) game, and can be computed in polynomial time in the size of their input games \cite{DBLP:conf/fossacs/HuthKP13,Huth15tr,HKP15}.
Static color compression $\scc$ is agnostic to the game graph and makes
$\Col G$ convex in
$\Nat$, e.g.\ $\Col G = \{0,2,3,6,7\}$ becomes $\{0,1,2,3\}$ where
nodes coloured with $2$ now have color $0$, nodes coloured $3$ now have
color $1$ and so forth. 
Priority propagation $\pp$ is informed by the
game graph. At node $v$, let $p(v)$ denote $\min(\max c(v\nav E),\max
c(E\nav v))$ where $c(Y) = \{c(y)\mid y\in Y\}$; if there is a node $v$
with $p(v) < c(v)$, one such node is selected by $\pp$ and the color at $v$ is
changed to $p(v)$; otherwise $\pp$ has no effect.

The monotone attractor for a node set $X$ of color $d$ in
$\Col G$ \cite{DBLP:conf/fossacs/HuthKP13} is defined as follows: 
it is the greatest set of nodes $Y_X$ in $G$ from which player $d\%
2$ can force to reach nodes in $X$ whilst only encountering nodes of
color $\geq d$ en route. A node set $X$ is a \emph{fatal}
attractor \cite{DBLP:conf/fossacs/HuthKP13} 
if it is contained in its monotone attractor $Y_X$,
and then all nodes in $X$ are won by player $d\%2$ in parity game $G$
\cite{DBLP:conf/fossacs/HuthKP13}. 
We write $\fa$ for the static analysis that returns a fatal
attractor (say by exploring colours in descending order) if $G$ has
one, and returns nothing otherwise. 
Another static analysis $\ari$ is based on the abstract Rabin
index of parity games \cite{HKP15}: for node $v$ with $c(v) > 1$, let
$c'_v$ be the maximal color of all cycles that go through node $v$ in
$G$; if there is a node $v$ with $c'_v < c(v)$, then $\ari$ chooses one such
node and changes the color at $v$ to $c'_v$; otherwise $\ari$ has no
effect on $G$. 
Finally, let $\gfa$ be a more general form of
partial solver $\fa$, based on the partial solver in \cite{Huth15tr}. 
For $\gfa$, all nodes in node set $X$ have color parity $p$
(not necessarily the same
color), and $Y_X$ is the greatest set of nodes from which player $p$ can
ensure that $X$ is reached such that the minimal color encountered en
route has parity $p$ \cite{Huth15tr}. Partial solver $\gfa$ returns a set $X$
contained in the corresponding $Y_X$, if there is such a pair
$(X, Y_X)$, and returns nothing otherwise.

\section{Composition of partial solvers}
\label{section:composition}
We now present our approach to expressing and composing partial
solvers with ease. Fundamental to this is the notion of a state $s$ which has
form
\[
(W_0,W_1,\rho,G',G)
\]

\noindent and models an intermediate state of computation within an
implicit composition context. Below, we write $s.G$ and so forth to refer to
such components of state $s$, write $s.V$, $s.E$ etc.\ for the
components of game $s.G$, and similarly for game $s.G'$. We may elide
prefix ``$s.$'' if state $s$ is clear from context.
The original input game is $s.G$ and parity game $s.G'$ is the
\emph{continuation} game that
still needs to be solved; node sets $s.W_p$ for $p$ in $\{0,1\}$ model those nodes
in $s.G$ for which the winner is already decided as player $p$; for $v$ in $V'$, node
set $v\nav \rho$ for $\rho\subseteq V'\times V$ represents those nodes in $V$ that have the same (not yet known) winner
in $s.G$ as $v$ has in $s'.G'$; and the winning
regions $\win p{s.G}$ of $s.G$ are the union of $s.W_p$ and the image of
the winning region $\win p{s.G'}$ under relation $\rho$. A state
models \emph{configurations} of partial solver computations, where
\[
(\emptyset,\emptyset,\Delta_{V_G},G,G)
\]

\noindent is a natural initial
configuration with $\Delta_{V_G} = \{(v,v)\mid v\in V_G\}$, and the
more general configurations model composition contexts.

We write $\Sigma$ for the set of all states, let the rank of $s$ be
the rank of $s.G'$, and define a partial order $\leq$ on states by
\[
s'\leq s\hbox{\ iff\ ($s=s'$ or $\rank
  {s'} < \rank s$)}
\]

\noindent Note that $(\Sigma,\leq)$ satisfies the descending chain condition, where the length of any descending
chain starting in $s$ is polynomial in $\rank {s.G'}$. 

A partial solver is a terminating algorithm $A$ whose semantics $f$ is a state
transformer of type $\Sigma\to \Sigma$ and satisfies, for all $s$ in
$\Sigma$, the following: the input game $s.G$ won't change under $f$,
$f$ strictly decreases the rank or won't change the state, and $f$ preserves winners of nodes whose winners have already been decided.

\begin{definition}
Let $\mathcal P$ be the set of partial solvers that run in polynomial time in rank of $s.G$.
\end{definition}

By abuse of language, we sometimes refer to functions $f$ as partial
solvers but context will determine which algorithms they denote.

Refinement for state transformers
$f\leq g$ is defined as
\(f\leq g\hbox{ iff }\forall s\in\Sigma\colon f(s)\leq g(s)\).
Then $f$ refines $g$, and so any partial solver with
semantics $f$ refines any partial solver with semantics $g$. 
We write $\Sigma_f$ for the set of \emph{residual games} of partial solver
    $f$, which $f$ cannot simplify.

We now formally present the five analyses from
Section~\ref{section:background} in this setting: Static color
compression $\scc$ maps a state $s=(W_0,W_1,\rho,G',G)$ to $s'$ which
is $s$ except that $s.G'$ may
change to reflect the compressed, convex color set. Priority
propagation $\pp$ also may only change $s.G'$ such that the color of
at most one node in $s.G'$ is decreased and all other aspects of
$s.G'$ remain the same in $s'$. For fatal attractor detection $\fa$, suppose it
detects a fatal attractor $X$ won by player $p$ in $G$. We set $Z
= \attr[p]{s.G'}{X}$. Partial solver $\fa$ then transforms state $s$
into (assuming $p=0$ without loss of generality):
\(s' = (W_0\cup Z\nav\rho, W_1,\rho',G'\setminus Z,G)\).
where $\rho'$ is the restriction of $\rho$ from domain $V'$
to $V'\setminus Z$, and $G'\setminus Z$ is parity game $G'$ restricted to node set $V'\setminus Z$ (which eliminates all incoming and outgoing edges of $Z$ as well). Next, consider static analysis $\ari$. If there is
no node $v$ in $s.G'$ with $c'_v < c(v)$, then $\ari(s) = s$. Otherwise,
some such node is chosen and $s'$ equals $s$ except that $s'.G'$
reduces the color at node $v$ to $c'_v$ in $s.G'$. The behaviour of $\gfa$
is the same as for $\fa$ above except that the manner in
which such a node set $X$ is computed differs \cite{Huth15tr}, e.g.\ colours of nodes
in $X$ may vary. 
We summarise the above discussion:
\begin{lemma}
\label{lemma:oldinp}
The partial solvers $\scc$, $\pp$, $\fa$, $\ari$, and $\gfa$ have semantics $\Sigma\to\Sigma$ and are in $\mathcal P$.
\end{lemma}

We are interested in sequential iterations of partial solvers that revert
control to the first solver in the sequence as soon as state rank decreases:
Let $f_1,\dots, f_k$ be elements of $\mathcal P$ with $k\geq 1$. Let,
for each $s$ in $\Sigma$, set $M_s$ be $\{ i\mid 1\leq i\leq k, \rank
{f_i(s)} < \rank s\}$. Then $\while {f_1, \dots, f_k}(s)$ is defined
as $s$ if $M_s=\emptyset$ and as $\while {f_1, \dots,
  f_k}(f_{j_m}(s))$ otherwise where $j_m = \min (M_s)$.
It is not hard to show that this defines a family of
operators on $\mathcal P$:
\begin{lemma}
\label{lemma:iterationinp}
For $k\geq 1$, operator $\lambda (f_1,\dots, f_k) \,\while {f_1,
  \dots, f_k}$ 
has type $\mathcal P^k\to \mathcal P$.
\end{lemma}

For a partial solver $g=\while {f_1, \dots, f_k}$,
we have $\Sigma_g = \bigcap_{i=1}^k \Sigma_{f_i}$. In particular,
$\Sigma_g$ is invariant under permuting
the order of the $f_i$ in $g$. 
Operator $\while {\cdot}$ supports our data-driven approach
to refinement as follows: given $g_0 = \while {f_1, \dots, f_k}$ we study
games $s.G'$ with $\while {f_1, \dots, f_k}(s) = s$ to manually learn
a new static analysis $f_{k+1}$ with $\while {f_1, \dots,
  f_k,f_{k+1}}(s) \not= s$, 
and then similarly consider $g_1 = \while {f_1,
  \dots, f_k,f_{k+1}}$ on the set of states $\Sigma _{g_0}$ for
further refinement. These are refinements since 
\(\while {f_1, \dots, f_k}\geq \while {f_1, \dots,
  f_k,f_{k+1}}\)
for all $k\geq 1$ and all partial solvers
$f_1,\dots,f_{k+1}$.  The partial solvers in \cite{DBLP:conf/fossacs/HuthKP13,DBLP:journals/corr/HuthKP14,Huth15tr} could not completely solve all 1-player games. We show that such completeness is achievable by the interaction of such partial solvers with $\ari$ and $\scc'$~--~a variant of $\scc$ that statically compresses the color set of each SCC in a parity game \emph{separately}: if $C$ is such a SCC with set of colours $\mathcal C$, then $\scc'$ makes $\mathcal C$ convex in $\Nat$ and recolours the SCC $C$ accordingly. This also illustrates how we may reason about states in $\Sigma_g$:
\begin{theorem}
\label{theorem:oneplayer}
Let $g= \while {f_1, \dots, f_k}$ be in $\mathcal P$ with
$\{\scc', \ari, \fa\}$ contained in $\{f_1,\dots, f_k\}$. Then there is
no $s$ in
$\Sigma_g$ for which $s.G'$ is a 1-player game.
\end{theorem}

Operator $\lifted {f}$ transforms a partial solver $f$ into a second-order
version that tests consequences of edge removals on residual games of
$f$. For game $G$ with edge relation $E$, this uses derived games:
\(G_{(v,w)}\) equals \(G\) except where $v\nav E$ is now $\{(v,w)\}$; and
\(G\setminus (v,w)\) equals \(G\) except where $(v,w)$ is removed from $E$. By abuse of notation, we write $s\setminus (v,w)$ for a state that equals state $s$ except that $(s\setminus (v,w)).G'$ equals $G'\setminus (v,w)$.
That is to say, $G_{(v,w)}$ removes from $E$ all edges
$(v,w')$ with $w\not=w'$, whereas $G\setminus (v,w)$ removes from $G$
the edge $(v,w)$.  The game \(G\setminus (v,w)\) will not introduce
deadlocks as it will only be called on nodes $v$ with $\size {v\nav E} >
1$. We also require notation for initial calling contexts of partial solvers:
\[
\call f(G) = (f(s).W_0,f(s).W_1)
\]

\noindent where $s$ equals $(\emptyset,\emptyset,\Delta_{V_G},G,G)$.
Expression $\call f(G)$ extracts the respective set of nodes that
$f$ can decide to be won by each player, when run in an initial
configuration for $G$. Operator $\lifted {f}$, in
Figure~\ref{fig:lifted}, tests whether the commitment to edge
$(v,w)$ in $G_{(v,w)}$ turns a residual state of $f$ into one that it
not residual, and this will allow it to simplify $G$ to either
$G_{(v,w)}$ or $G\setminus (v,w)$. Thus, $\lifted {\cdot}$ either leaves a
state unchanged or removes from $s.G'$ at least one edge.
We use $\lifted {\cdot}$ for defining, for all $f$ in $\mathcal P$, function $\lift f\colon \Sigma\to\Sigma$ through
\[
\lift f = \while {f, \lifted f}
\]
\begin{figure}
$li\!f\!ted(f)(s)\ \{$

\ \ $let\ H = (V^*,V^*_0,V^*_1,E^*,c^*)\ be\ s.G';$

\ \ $f\!or\ (v\ in\ V^*\ such\ that\ \size{v\nav E^*} > 1)\ \{$

\ \ \ \ $p = owner(v);$

\ \ \ \ $f\!or\ (w\ in\ v\nav E^*)\ \{$

\ \ \ \ \ \ $let\ (U_0,U_1) = \call f(H_{(v,w)});$

\ \ \ \ \ \ $i\!f\ (v\ in\ U_p)\ \{$

\ \ \ \ \ \ \ \ $return\ (s.W_0,s.W_1,s.\rho,H_{(v,w)}, s.G);$

\ \ \ \ \ \ $elsei\!f\ (v\ in\ U_{1-p})\ \{$

\ \ \ \ \ \ \ \ $return\ (s.W_0,s.W_1,s.\rho,H\setminus (v,w), s.G)$;

\ \ \ \ \ \ $\}$

\ \ \ \ $\}$

\ \ $\}$

\ \ $return\ s;$

$\}$
\caption{Pseudo-code for function $\lifted {\cdot}$ with dependent
  type $\prod_{f\colon \mathcal P}\, (\Sigma_f\to \Sigma)$: for
partial solver $A$ in $\mathcal P$ with semantics $f$, it renders a
partial solver $\lifted A$ in $\mathcal P$ with semantics $\lifted
f\colon \Sigma_f\to \Sigma$ by testing effects of edge removals on
running $A$\label{fig:lifted}}
\end{figure}

\noindent Note that $\lift f$ now has domain $\Sigma$ as the semantics
of $\while {\cdot}$ enforces that $\lifted f$ is only reached with
input from $\Sigma_f$. Let algorithm $A$ have semantics $f$; we write
$\lifted A$ for the algorithm obtained
from the pseudo-code for $\lifted f$ in Figure~\ref{fig:lifted} when all applications of $f$ are
implemented by $A$. Then $\lift A$ denotes $\while {A, \lifted A}$.
\begin{lemma}
\label{lemma:lift}
If $A$ in $\mathcal P$ has semantics $f$, then $\lift A$ is in
$\mathcal P$ and has semantics $\lift f$.
\end{lemma}

\noindent Of course, we may appeal to Lemma~\ref{lemma:lift}
repeatedly to define higher-order versions $\lift {\lift A}$ and so
forth for algorithms $A$ in $\mathcal P$ with semantics $f$, which are
all in $\mathcal P$ by virtue of this lemma. Next, we use these
operators for data-driven refinement.

\section{Data-driven refinement}
\label{section:refinement}
Let us first consider partial solver
\[
\psolone = \while {\scc, \pp, \fa, \ari, \gfa}
\]

\noindent Based on the semantics of $\while {f_1,\dots,f_k}$, we may assume that the input domain of each $f_j$ with $j > 1$ equals $\bigcap _{i=1}^{j-1} \Sigma_{f_i}$. In particular, if some partial solver $f_j$ requires that its input games have no fatal attractors, this is guaranteed by having $f_l = \fa$ for some $l < j$. We will also that
a new analysis $f_{k+1}$ (which may be more expensive, say) is only ever called in the refinement $\while{f_1,\dots,f_{k+1}}$ on states that are residual for $\while{f_1,\dots,f_k}$.

Some static analyses below will merge a set of nodes $X$ to a sole node owned by player $p$ and of color $d$. This merge operation can be defined generically:
\begin{definition}
Let $s$ be a state, $X\subseteq V'$ with $\size X\geq 2$ and $X\nav E'\setminus X\not=\emptyset$.
Let $p$ be a player, $d$ a color, and $z\not\in s.V'$. Then tuple $merge(s,X,p,d,z)$ denotes
\begin{equation}
\label{equ:mergedstate}
(s.W_0,s.W_1,merge(\rho,X,z), merge(s.G',X,p,d,z),G)
\end{equation}
\noindent  where the parity game $merge(s.G',X,p,d,z)$ is defined as $(V^*,V^*_0,V^*_1,E^*,c^*)$ with
\begin{eqnarray*}
V^*_{1-p} &=& V'_{1-p} \setminus X\\
V^*_p &=& (V'_p \setminus X)\cup \{z\}\\
E^* &=& (E'\setminus X\times X)\cup ((E'\nav X\setminus X)\times \{z\})
\cup (\{z\}\times (X\nav E'\setminus X))
\end{eqnarray*}

\noindent and $c^*(v) = c'(v)$ for all $v\not=z$ whereas $c^*(z) = d$.
Relation $merge(\rho,X,z)$ is $(\rho\setminus X\times s.V')\cup \{ (z,w)\mid w\in X\nav\rho\}$.
\end{definition}

Whenever we invoke the above merge method, we need to ensure that the resulting
tuple is an actual state.
The parity game $merge(s.G',X,p,d,z)$ has no dead-ends: this is so since $z$ has at least one outgoing edge, which is guaranteed by the fact that $(x,v)$ is in $s.E'$ for some $x$ in $X$ and some $v$ in $s.V'\setminus X$. 
Next, we present two static analyses that use this merging.

\paragraph{{\bf Sole successor node merging: $\mss$.}}
An inspection of residual games for $\psolone$ identifies a method
$\mss$ for merging two nodes, so we set
\[
\psoltwo = \while {\scc, \pp, \fa, \ari, \gfa, \mss}
\]

\begin{figure}
\centering
\includegraphics[scale=\gamesizeone]{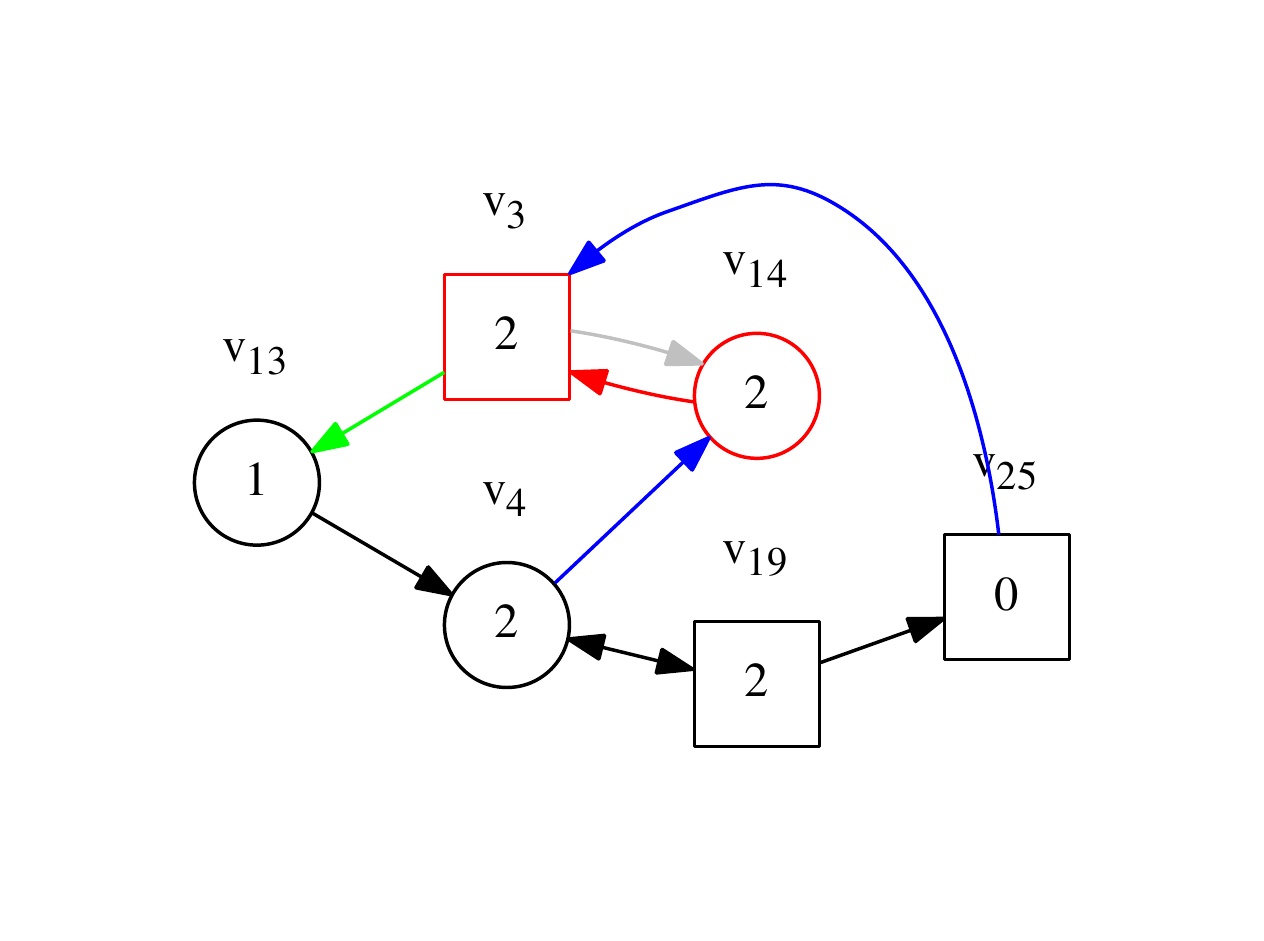} 
\includegraphics[scale=\gamesizeone]{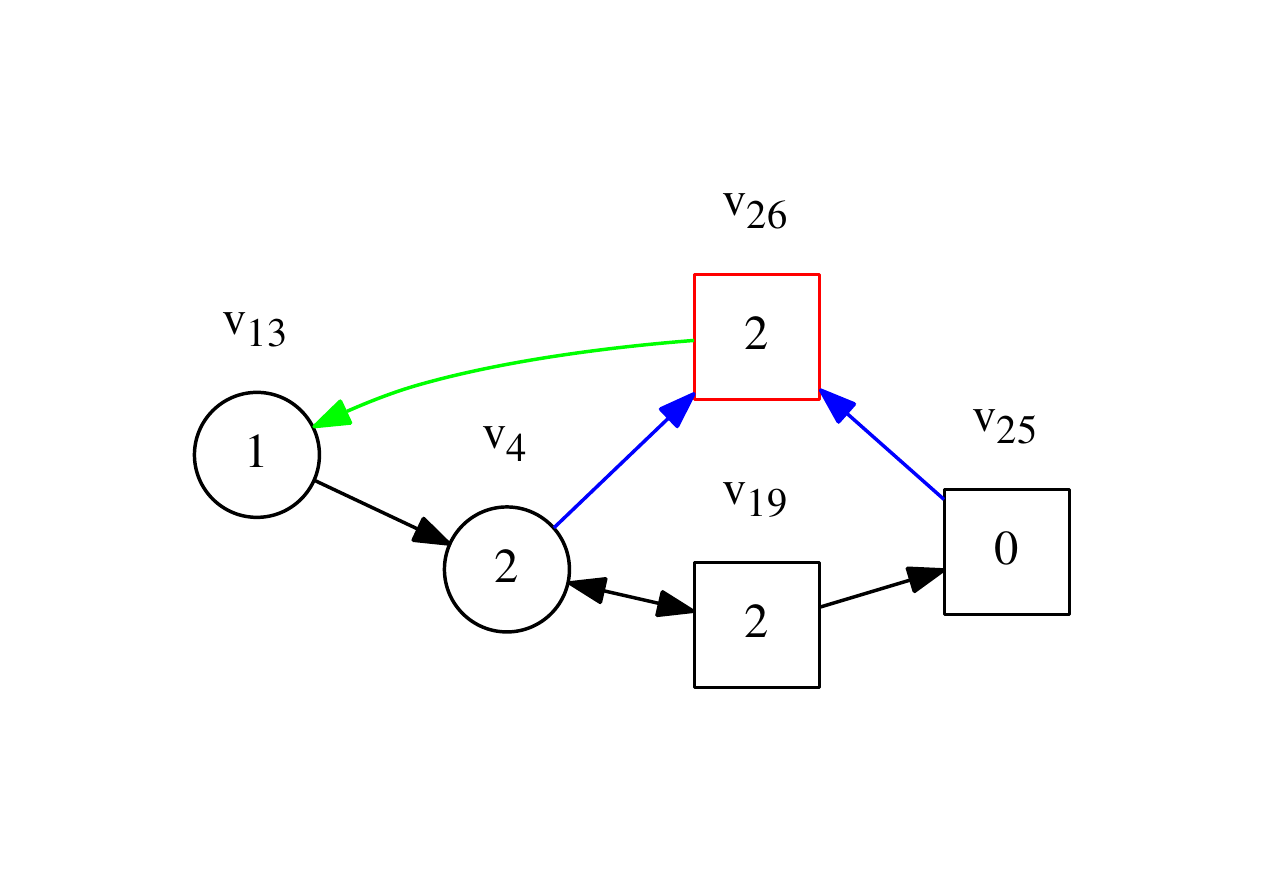} 
\caption{Left: Residual game $s.G'$ for $\psolone$. Right: game $\mss(s).G'$  obtained from the call $merge(s, \{v_{14},v_3\},owner(v_3), c(v_3),v_{26})$\label{fig:mss}}
\end{figure}

\noindent To see how
$\mss$ works, let $s$ be a state in $\Sigma$. Suppose that there are
two nodes $v$ and $w$ in $s.G'$ such that $v\nav E' = \{w\}$, $w\nav E'\not\subseteq \{v,w\}$, and the color of $v$ in $s.G'$ is not smaller than that of $w$. Choose some $z$ not in the node set of $s.G'$. Then
\(\mss(s) = merge(s, \{v,w\},owner(w), c(w),z).\)
As $w\nav E'$ contains a node not in the merge set
$\{v,w\}$, state $\mss(s)$ is well defined. If there are no such nodes $v$ and $w$, we set $\mss(s) = s$.
Figure~\ref{fig:mss} shows a residual game for $\psolone$ and the
effect of $\mss$ on it: node $v_{14}$ is $v$, node $v_3$ is $w$, the
owner of $w$ is player $1$, $w$ has color $2$, and $z$ is $v_{26}$.

\begin{theorem}
\label{theorem:mss}
The static analysis $\mss$ is in $\mathcal P$.
\end{theorem}

\paragraph{{\bf Merging SCCs: $\mscc$.}}
The study of residual games for $\psoltwo$ introduces more complex
methods for merging nodes. We will only describe one of these next,
static analysis $\mscc$ which operates on states residual for $\fa$ and 
attempts to merge an SCC in a sub-game
of the residual game. For state $s$, this analysis checks whether
there is some color $d$ such that the following can be realised: Let
$H=(V'[Z],E'\cap Z\times Z)$ be the game graph that restricts the game
graph of $s.G'$ to $Z = \{w\in V'_{1-p}\mid c(w) \geq d\}$, the set of
all nodes $w$ owned by player $1-p$ and of color $\geq d$ in $s.G'$
where $p= d\%2$. Suppose there is an SCC $C$ in $H$ and a subset
$X\subseteq C$ with $\size X > 1$ such that all elements in $X$ have
color $d$ and where $X\nav E'\cap (V'\setminus X)\not=\emptyset$ in $G'$. The latter implies
\(\mscc(s) = merge(s, X, 1-p, d,z)\)
is well defined: since $X\nav E'\cap (V'\setminus
X)\not=\emptyset$, the parity game $\mscc(s).G'$ contains no dead-ends. If there is no such color $d$ with corresponding $H$ and $X$, we set $\mscc(s)$ equal to $s$. This defines a refined partial solver
\[
\psolthree = \while {\scc, \pp, \fa, \ari, \gfa, \mss, \mscc}
\]

\begin{figure}
\centering
\includegraphics[scale=\gamesizetwo]{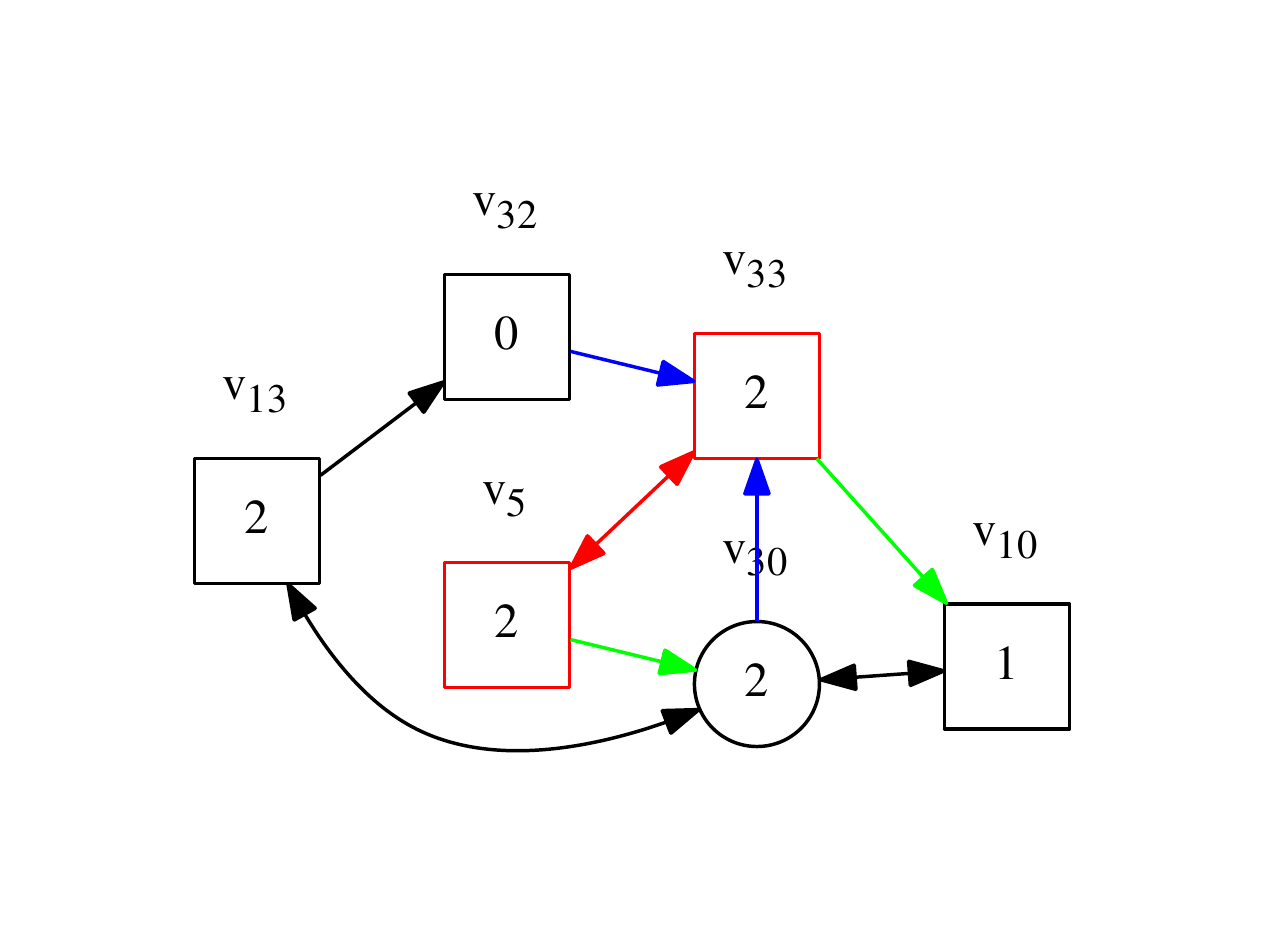} 
\includegraphics[scale=\gamesizetwo]{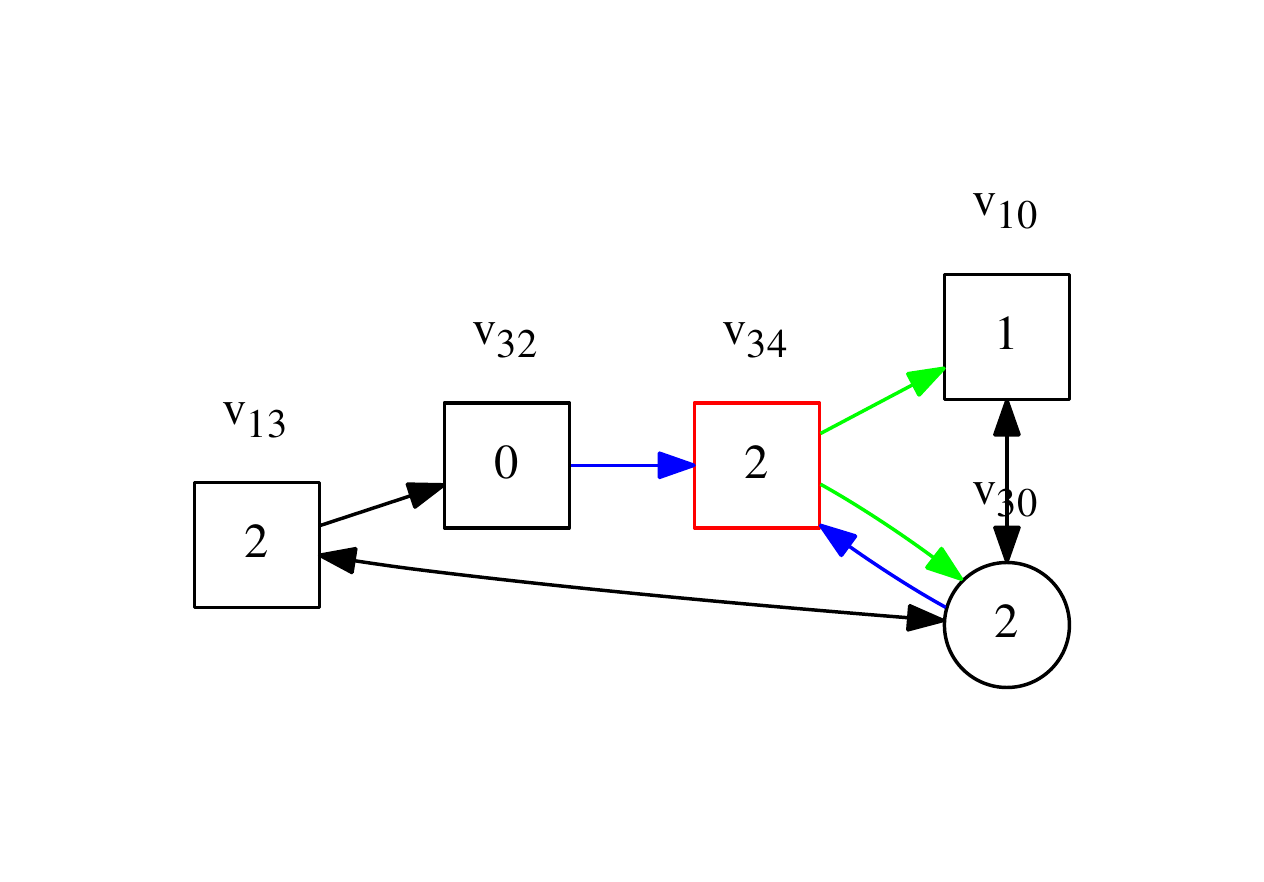} 
\caption{Left: Residual game $s.G'$ for $\psoltwo$. Right: game $\mscc(s).G'$  obtained from the call $merge(s, \{v_5,v_{33}\},1-0, 2,v_{34})$\label{fig:mscc}}
\end{figure}

\noindent Figure~\ref{fig:mscc} shows a residual game for $\psoltwo$ and the effect of $\mscc$ on it: $d$ is $2$, $p$ is $0$, node set $X$ is $\{v_5,v_{33}\}$, and $z$ is $v_{34}$.

The soundness proof for this analysis is pretty straightforward: first
we show that the same player indeed wins all nodes in $X$, and then we
show that the merged version of the continuation game has the same
winning region modulo $\rho$.

\begin{theorem}
\label{theorem:mscc}
The static analysis $\mscc$ is in $\mathcal P$ with domain $\Sigma_{\fa}$.
\end{theorem}

\paragraph{{\bf Edge removal based on conditional fatal attractors: $\erfa$.}}
The residual games of $\psolthree$ led us to studying edge removal methods
for states in $\Sigma_{\fa}$. We discovered static analysis $\erfa$
which works as follows for any $s$ in $\Sigma_{\fa}$: 
If there is an edge $(v,w)$ in $s.G'$ such that $s.G'_{(v,w)}$ has a fatal attractor, then $\erfa$ choses one such edge and sets
\(\erfa(s) = s\setminus (v,w)\), i.e.\ $\erfa$ removes edge $(v,w)$ from $s.G'$.
The intuition is that any fatal attractor that would appear in  $s.G'_{(v,w)}$ would have to be a fatal attractor for player $1-owner(v)$, since $s$ is in $\Sigma_{\fa}$. Therefore, we may remove the edge $(v,w)$ from $s.G'$ as choosing this edge would lead player $owner(v)$ to lose that node.
Otherwise, if no such edge exists, $\erfa(s)$ equals $s$. For refined partial solver
\[
\psolfour = \while {\scc, \pp, \fa, \ari, \gfa, \mss, \mscc,\erfa}
\]

\begin{figure}
\centering
\includegraphics[scale=\gamesizethree]{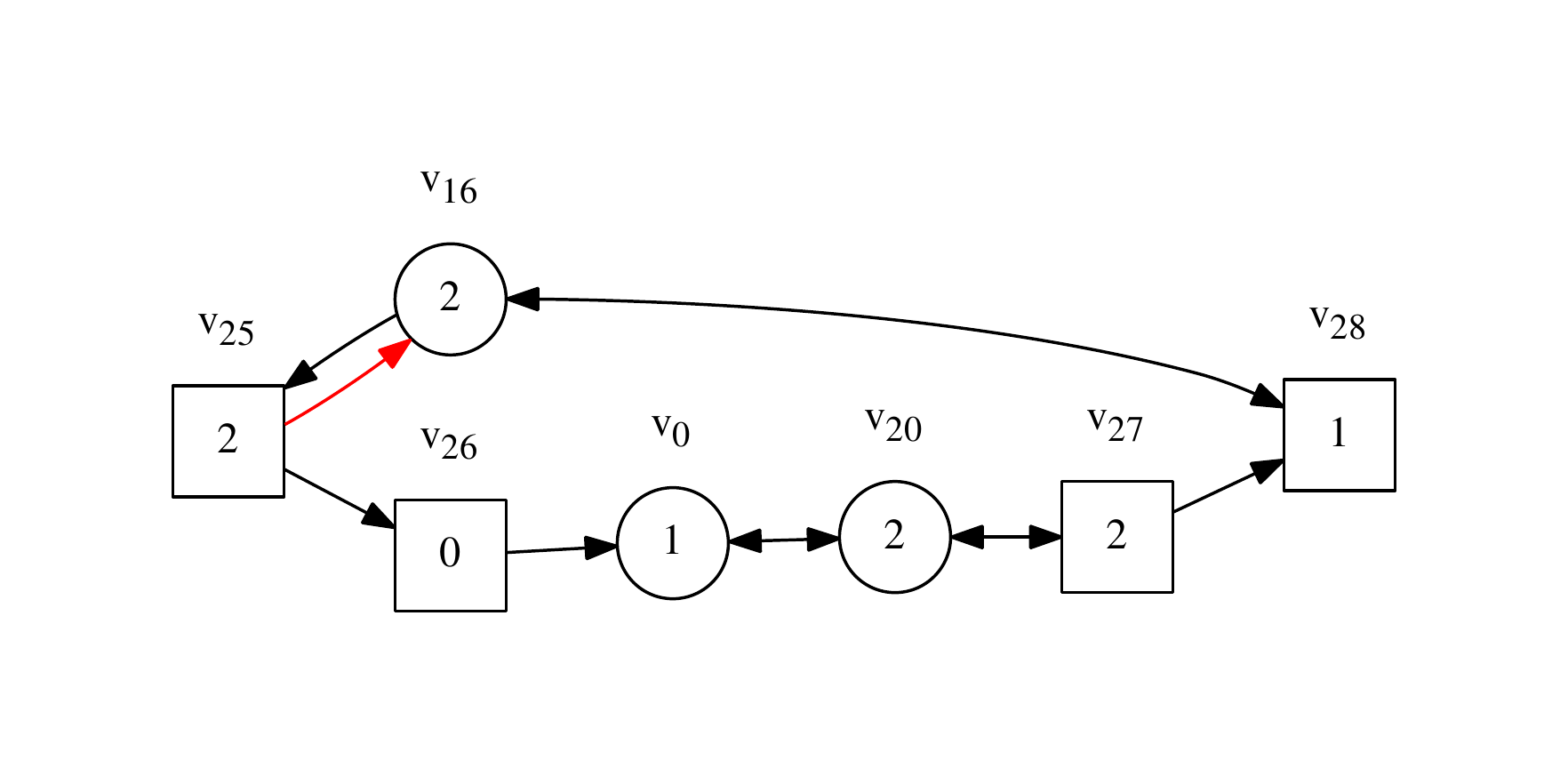} 
\includegraphics[scale=\gamesizethree]{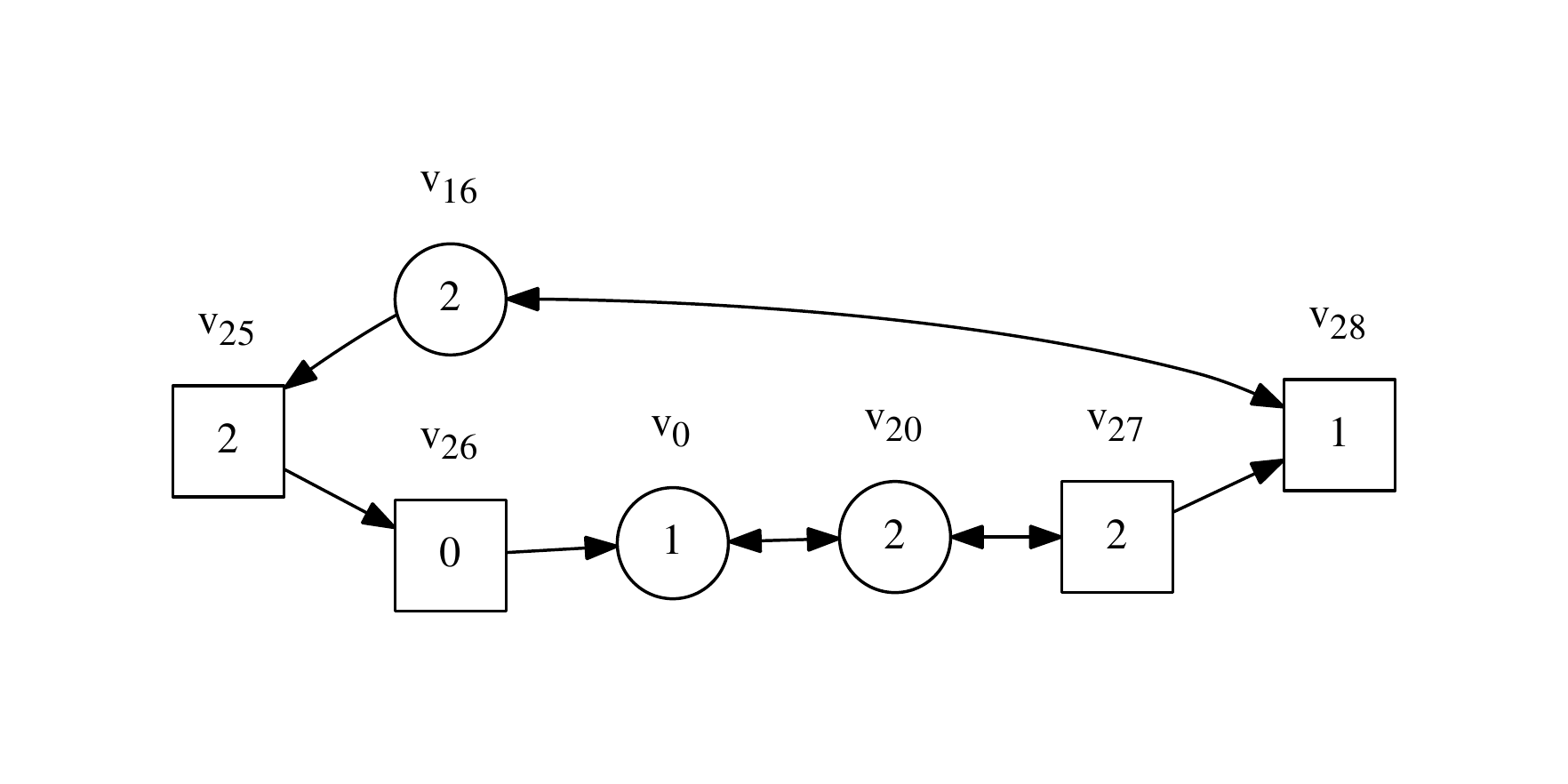} 
\caption{Left: Residual game $s.G'$ for $\psolthree$. Right: game $\erfa(s).G'$  which removes edge $(v_{25},v_{16})$ since $s.G'_{(v_{25},v_{16})}$ contains $\{v_{25},v_{16}\}$ as a fatal attractor for color $0$\label{fig:erfagame}}
\end{figure}

\noindent Figure~\ref{fig:erfagame} shows $s.G'$ for some $s$ in $\Sigma_{\psolthree}$ and the effect of $\erfa$ on it: $d$ is $2$, $p$ is $0$, set $X$ is $\{v_5,v_{33}\}$, and $z$ is $v_{34}$.

\begin{theorem}
\label{theorem:erfa}
The static analysis $\erfa$ is in $\mathcal P$ with domain $\Sigma_{\fa}$.
\end{theorem}

\paragraph{{\bf Edge removal based on shared descendant: $\ersd$.}}
Residual games for partial solver $\psolfour$ suggested to us the following static analysis $\ersd$, which removes an edge based on a shared descendant. This checks, for $s$ in $\Sigma$, whether there are three different nodes $v,w,z$ in $s.G'$, an edge $(v,w)$ in $s.G'$, $p$ in $\{0,1\}$, and two colours $c_v$ and $c_w$ in $\Col {s.G'}$ (not necessarily at $v$ or $w$) with $c_v\preforder p c_w$ such that:
\begin{compactitem}
\item there is a path $P_{vz}$ of color $c_v$
  from node $v$ to $z$ in $s.G'$ such that all nodes on $P_{vz}$ are in $V_p$ or have only one outgoing edge in $s.G'$, and
\item  there is a path $P_{wz}$ of color $c_w$
  from node $w$ to $z$ in $s.G'$ such that all nodes on $P_{wz}$ are in $V_{1-p}$ or have only one outgoing edge in $s.G'$.
\end{compactitem}

\begin{figure}
\centering
\includegraphics[scale=\gamesizefour]{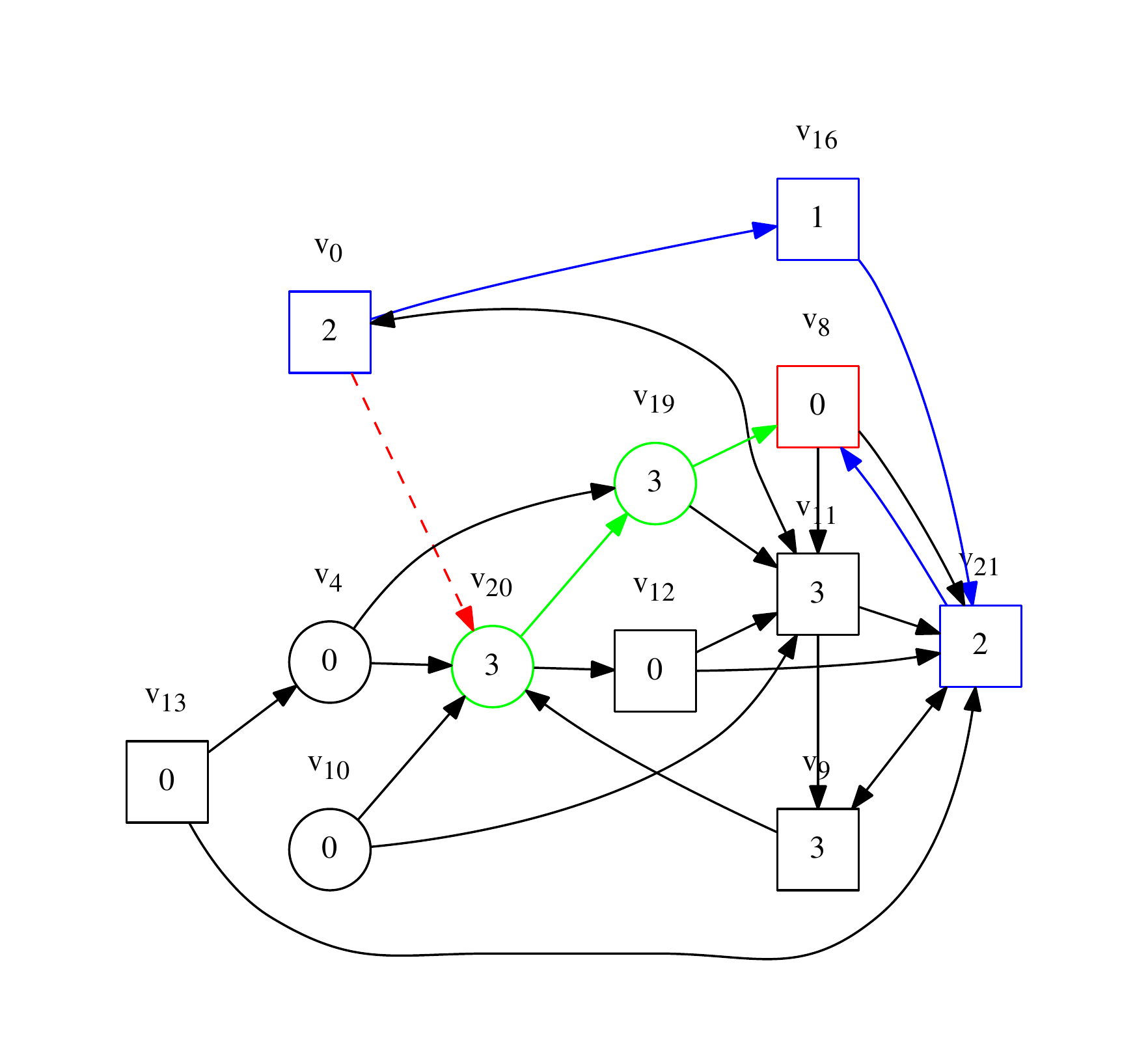} 
\caption{Left: Residual game $s.G'$ for $\psolfour$. Right: game $\ersd(s).G'$  which removes edge $(v_{0},v_{20})$ since it has two control paths that meet the criteria for $\ersd$\label{fig:ersd}}
\end{figure}

\noindent If there are such data, $\ersd$ chooses one such edge $(v,w)$ and sets
\(\ersd(s) = s\setminus (v,w)\), i.e.\ edge $(v,w)$ is removed from $s.G'$.
The intuition is that this only requires an argument when player $p$
wins $v$ in $s.G'$ with a winning strategy that moves from $v$ to $w$:
then we can employ a \emph{dominance} argument based on $\preforder p$
as indicated below.
Otherwise, if no such edge exists, $\ersd(s)$ equals $s$. 
Figure~\ref{fig:ersd} shows the effect of $\ersd$ on a residual game for $\psolfour$: $v$ is $v_0$, $w$ is $v_{20}$, $p$ is $1$, $z$ is $v_8$, the path $P_{vz}$ (blue, via $v_{16}$ and $v_{21}$) has color $c_v = 0$, and the path $P_{wz}$ (green, via $v_{19}$) has color $c_w=0$.
This yields a refined partial solver
\[
\psolfive = \while {\scc, \pp, \fa, \ari, \gfa, \mss, \mscc,\erfa, \ersd}
\]

\noindent The proof of the correctness of $\ersd$ exploits that removing an edge
$(v,w)$ where $v$ is in $V_p$ cannot increase the winning region of
player $p$. Therefore, it will suffice to show that this does not
\emph{decrease} the winning region of player $p$.
Only the case when
$v$ is won by player $p$ with a strategy that moves from $v$ to $w$ is
of real interest: then,  any winning strategy for player $p$ at node $v$ in $s.G'$ can be replaced with a winning strategy in $s.G'\setminus (v,w)$ that moves from node $v$ along the path $P_{vz}$. In detail,
we then use this strategy $\tau$ and the path
$P_{vz}$ to define a new
strategy $\gamma$ with finite memory for that player on the new game
$G'' = s.G'\setminus (v,w)$. We then show that this new strategy
$\gamma$ is winning in game $G''$ on the old winning region of $s.G'$,
by showing that each infinite play in game $G''$ conformant
with the new strategy $\gamma$ determines an infinite play in
game $s.G'$ that is conformant with the (winning) strategy $\tau$,
such that the outcome for player $p$ of the infinite play in 
game $G''$ is better or equal with respect to $\preforder p$ to the
outcome of the infinite play in game $s.G'$. This ensures that
player $p$ wins the infinite play in the new game $G''$, as he does
win the infinite play in game $s.G'$.

\begin{theorem}
\label{theorem:ersd}
The static analysis $\ersd$ is in $\mathcal P$.
\end{theorem}

\section{Experimental results}
\label{section:experiments}
Our approach and its implementation in Python do not compute winning
strategies since soundness proofs for some partial solvers require
finite memory;
related to that, in \cite{Huth15tr} it was noted that
the partial solver $\psolc$, to which $\while {\gfa}$ in our paper is similar, may
require finite memory.
We use 
PGSolver \cite{Friedman09,Friedmann10}
as a test oracle to validate that our implementations of partial
solvers are sound, i.e.\ that they never misclassify the winner of a node of an
input game. 

\paragraph{{\bf Experiments on structured benchmarks.}}
We ran $\psolone$ on Keiren's comprehensive benchmark
suite \cite{keiren15} on a HP EliteDesk 800 G1 TWR with RAM 16GB and an
Intel Core i7-4770 3.40GHz. For efficiency reasons, we ran $\psolone$ over
all games in that suite whose textual representation was less than
200KB. This suite contains the PGSolver benchmarks as well; however,
for some of the latter types Keiren's suite only contains games whose
textual representation is larger than 200KB; for these types we thus
used PGSolver itself to generate such test games.
We refer to \cite{TR} for the full list of these games.
In this manner, we tested
481 games~--~some of which with more than 10,000 nodes. Both $\psolone$ and our implementation of Zielonka's algorithm solved 464 of these games completely and agreed on those solutions. For the remaining 17 games, an exception was raised (stack overflow or a timeout of 60 seconds) for at least one of $\psolone$ or our implementation of Zielonka's algorithm. Our version of Zielonka's algorithm was also extensively tested against the PGSolver command $pgsolver\ -global\ recursive$, justifying its use in validation testing. That use allowed us to unit test more efficiently, as our pipe from Python to PGSolver input was rather slow.

\paragraph{{\bf Random games used.}}
We used a standard type of random game \cite{Friedmann10} with configuration $xx$-$yy$-$aa$-$bb$, which has
$xx$ nodes whose ownership is determined uniformly at random, $yy$
colours where colours of nodes are independently and uniformly drawn
from set $\{0,1,\dots, yy\}$, and where for each node $v$ the set
$v\nav E$ has at least $aa$ and at most $bb$ elements; the cardinality
of $v\nav E$
is determined for each node independently and uniformly at random.

\paragraph{{\bf Unit testing for our implementation of solvers.}}
For each of the four new analyses of Section~\ref{section:refinement}, we generated a stream of random games and applied the analysis to each game as often as it would result in state changes. For each state change, we tested whether the winning regions (modulo potential node merging via $\rho)$ won't change. Specifically, we generated 100,000 such tests for each analysis. For $\erfa$, we used configuration 60-30-2-3 and 100,395 games in $\Sigma_{\fa}$ to generate that many tests. For $\ersd$, we used configuration 60-30-2-3 with 24,081 games, for $\mss$ we took configuration 60-30-1-3 and  100,140 games in $\Sigma_{\fa}$, and for $\mscc$ we had configuration 60-30-1-3 with 1,885,423 games.
Note that such tests may generate fewer games than test cases, if the analysis can be applied repeatedly on continuation games. But we may have to generate more games than tests, which was the case for analyses that require states from $\Sigma_{\fa}$.
\hide{%
\begin{figure}
{\small
\begin{verbatim}
### unit_er_fa for configuration 60-30-2-3###
analyses_max = 100 000, total number of games : 100395 (in Sigma_fa)

### unit_er_sd for configuration 60-30-2-3 ###
analyses_max = 1 000 000, total number of games : 24081

### unit_m_ss for configuration 60-30-1-3###
analyses_max = 100 000, total number of games : 100140 (in Sigma_fa)

### unit_m_scc for configuration 60-30-1-3 ###
analyses_max = 100 000, total number of games : 1885423
\end{verbatim}
}
\caption{Unit tests for analyses $\erfa$, $\ersd$, $\mss$, and $\mscc$\label{fig:unittestnew}. All tests passed. {\tt analyses\_max} records how often an analysis changed a state and passed the corresponding test, and {\tt Total number of games} lists how many games had to be generated to create that many tests. For $\mss$ and $\mscc$ we used a different configuration type to help with getting that many test cases}
\end{figure}
}

In addition, we did unit testing of partial solvers $\psolone$ through to
  $\psolfive$: we generated 10 million games of type 50-25-2-4 as a
  test harness; these partial solvers $\psolone$ through to
  $\psolfive$ never misclassified a node for
  all of these games, based on the regression test with PGSolver as
  described above. Here we also unit tested that these are refinements: $\psolone\geq \psoltwo\geq \psolthree\geq \psolfour\geq \psolfive$. This gave us high confidence that these
  implementations are correct. So we turned unit tests off in further
  experiments that explored billions of random games in search for
  residual games.

Finally, we unit tested $\lifted{\cdot}$ on 974 residual games that we found for $\psolfive$ and ran $\lifted{\psolfive}$ on those: for all of these games 
this call removed at least one edge (i.e.\ it reached the {\tt if} or {\tt elseif} branch in Figure~\ref{fig:lifted}) and it successfully tested that no winning regions changed. We did the same unit tests on 24,132 residual games for $\psolfour$ that we generated. For each of these games, the {\tt if} or {\tt elseif} branch was reached and the resulting game did not change winning regions.

\paragraph{{\bf Comparing effectiveness of new analyses.}}
We wanted to understand how often these four analyses can simplify
games. For this, we considered states in $\Sigma_{\fa}$ to create an
input common to all these analyses. We generated 100,000 states $s$ in
$\Sigma_{\fa}$ where $s.G'$ is the result of eliminating all fatal
attractors from a random game of configuration type 60-30-2-3. The
analyses simplified 99,596 such games for $\erfa$, 84,126 games for
$\ersd$, 80,327 for $\mss$, and 7,946 for $\mscc$. Then we did a
similar experiment for 25,360 residual games of a partial solver
similar to $\psolthree$, whose residual games are all in
$\Sigma_{\fa}\cap \Sigma_{\erfa}\cap \Sigma_{\mss}$: this confirmed that neither $\erfa$ nor $\mss$ simplified any of these games~-~~whereas $\ersd$ simplified 25,355 of these and $\mscc$ simplified 20,119 of these.

\paragraph{{\bf Experiments for data-driven refinement.}}
We conducted experiments to determine which random game configurations
$xx$-$yy$-$aa$-$bb$ are more prone to generating residual games for our partial solvers
above: when $bb$ equals $aa+2$ and $xx$ and $yy$ are fixed,
we noticed that $bb=2$ was most
effective at generating residual games whereas $aa\geq 5$ was very
ineffective. Fixing $aa$ and $bb$ and letting $yy$ be $xx$ or $xx/2$,
we noted that residual games occur more frequently as $xx$ increases
from $30$ to about $90$ but then occur less frequently again. Fixing
only $yy$, we noted that an increase beyond $15$ did not have much
effect.
These insights informed a large experiment in which we generated
$10,422, 420$ random games of type 50-25-2-3 in total~--~more than 10 million games~--~and recorded how
many residual games each of the five partial solvers had for these:
$32, 716$ for $\psolone$, $30, 631$ for $\psoltwo$, $19, 230$ for
$\psolthree$, $958$ for $\psolfour$, and only $136$ for
$\psolfive$. The latter $136$ residual games are completely solved by  $\lift{\psolfive}$.
This illustrates that each of the newly discovered
partial solvers leads to more effective refinements of existing ones.

\paragraph{{\bf Experiments for $\lift{\cdot}$.}}
We ran $\lift{\psolfive}$ on a range of random game configurations to
see whether we could find any non-empty residual games. We tested this
on games of varying configurations with node sizes ranging from 40 to
1000. \emph{All} of these games, totalling to 9,353,516,890 (over nine
billion games), were solved completely by $\lift{\psolfive}$;
specifically, we first ran $\psolfive$ on these games and invoked
$\lift{\psolfive}$ on all the non-empty residual games, which were
only in the order of thousands. This staging is justified as
$\psolfive$ is part of the interaction within $\lift{\psolfive}$.

\paragraph{{\bf Experiments on large games.}}
The implementation of our game generators and partial solvers is not optimized. For one, it may take too long to generate random games of one million nodes or more. For another, our partial solvers may not be able to solve such large games in a reasonable period of time, be it for random or structured games. Therefore, the insights reported above, including the effectiveness of our proposed data-driven approach to discovering new partial solvers, are limited in that they refer to parity games of small or medium size. However, we did mean to generate~--~within these performance constraints~--~some large games and determine whether $\psolfive$ would be able to solve them completely. The limited number of games that we managed to generate and test in this manner will only provide anecdotal evidence. But we do report such evidence here for sake of completeness.

We tested $\psolfive$ against the games from PGSolver's structured benchmark that are used in \cite{Friedman09}. We solved the {\sf Elevator} verification problem $G_n$ for $2 \leq n \leq 5$ and its variant $G_n'$ for $3 \leq n \leq 6$. We also solved the {\sf Tower of Hanoi} problem $T_n$ with $5 \leq n \leq 11$. 
The size of the corresponding games varied from $564$ nodes to $708,587$ nodes. The partial solver $\psolfive$ completely solved all of these structured games.

We also tested $\psolfive$ on large games from the remaining categories of Keiren's comprehensive benchmark suite, namely {\sf mlsolver}, {\sf equivchecking} and {\sf modelchecking}. We managed to generate $15$ games across these three types of benchmarks, where the size of these games varied from $35,234$ nodes to $1,081,474$ nodes. Partial solver $\psolfive$ completely solved all of these $15$ games.

Based on the experiments on random games that were performed in \cite{Friedman09}, we finally ran both $\psolfive$ and $\lifted{\psolfive}$ on random games with node sets of size either ten thousand, one-hundred thousand, or one million. Specifically, we managed to generate $35,699$ games with $10,000$ nodes ($32,054$ games in configuration $10000$-$100$-$2$-$4$ and $3645$ games in configuration $10000$-$1000$-$2$-$4$), $882$ games with $100,000$ nodes of configuration  $100000$-$100$-$2$-$4$, and $17$ games with $1,000,000$ nodes with configuration $1000000$-$yy$-$2$-$4$ where $yy$ equals $100$ for six games, $10$ for one game, $22$ for four games, and $50$ for six games. All of these $36,598$ games were completely solved by $\psolfive$, and so therefore also by $\lifted{\psolfive}$.

\section{Related work}
\label{section:related}
In \cite{Friedman09}, a pattern is proposed, implemented, and evaluated for how to solve parity games. This generic solver can be seen as a composition context of partial solvers (in our setting and terminology) in which all but one partial solver run in polynomial time, and where the latter is a complete solver that is only called when the partial solvers cannot progress on any terminal SCC of the parity game. The aim of this is to gain efficiency, and this was successfully demonstrated in \cite{Friedman09}. But the aim of our work here is to gain \emph{effectiveness} so that a composition context of partial solvers would never or very rarely have to call a complete solver.
In~\cite{Friedmann11}, it is shown that a variant of Zielonka's algorithm solves some classes of parity games in polynomial time, and an improved lower (exponential) bound is derived for solving all parity games with such recursive algorithms.
In \cite{DBLP:journals/eceasst/HuthPW09}, a function related to $\lift f$ is studied; using our terminology, it operates as follows: for $w\not= w'$
in $v\nav E$, if there is some node $z$ in $G$ such that $f$ detects a
different winner for node $z$ in the two games $G_{(v,w)}$ and
$G_{(v,w')}$, then node $v$ is won by player $owner(v)$ in parity game
$G$. It would be of interest to integrate this method into our
approach for experimental evaluation.
In \cite{DBLP:journals/corr/HuthKP14}, another function similar to $\lift f$ is investigated: apart from presentational differences (our work here uses states), the function in \cite{DBLP:journals/corr/HuthKP14} essentially omits the $i\!f$ part of code in Figure~\ref{fig:lifted} and its soundness proof had severe restrictions on the types of partial solvers that it may use as arguments.
In \cite{Huth15tr}, experiments compared the effectiveness of partial
solver \psolb  of \cite{DBLP:conf/fossacs/HuthKP13} 
(which is similar to our $\while {\fa}$) and \psolc (which is similar
to our $\while {\gfa}$): on random games, \psolc was more effective than
\psolb on games with higher edge density, but not at all more effective on
games with lower edge density.

\section{Discussion}
\label{section:discussion}
We also ran detailed experiments on residual games of some of the partial solvers $\psolone$ to $\psolfive$. Specifically, we studied structural features of their terminal SCCs. It appears that such SCCs have statistically significant structure. For example, we were unable to find a terminal SCC of a residual game that has two winners; however, we could then manually combine two such games to construct a residual terminal SCC in which both players win nodes. 

We implemented the partial solver $\ersd$ in a weaker version than that presented above: control paths only have nodes \emph{owned} by the controlling player. It may be possible to generalise the $\ersd$ specified in the paper such that node $z$ is reached in the alternating sense by the controlling player (on a tree rather than on a path), and always reached with the specified color.

Our approach to data-driven refinement of partial solvers worked well
since residual games were found within a reasonable amount of
time. But this method led to powerful partial
solvers for which we now genuinely struggle to find any residual games
by relying on standard random and non-random benchmarks. This may make
it harder to evaluate and improve such a partial solver.
Theorem~\ref{theorem:oneplayer}, however, suggests one form of
evaluation: to prove
mathematical properties of residual games that may also imply that
well known types of games are never residual for a given partial
solver. Partial solver $\psolone$, e.g., completely solves all
B\"uchi games, as $\psolb$ in \cite{DBLP:conf/fossacs/HuthKP13} does that.

Our paper focussed on \emph{effectiveness}: the ability of a partial
solver to completely solve a game in polynomial time. Our approach
can also facilitate the study of the \emph{efficiency} of composed
partial solvers, for example by choosing the order of arguments in
$\while{\cdot}$.

\section{Conclusions}
\label{section:conclusion}
There are many heuristics for solving or preprocessing parity games,
potentially decreasing the complexity of a parity game by reducing
some of its colours, by removing some of its edges, or by removing some
of its nodes (whose winners would then be known). Such methods are
sound as they do not alter the winning regions of the resulting parity
game. We developed here an approach to composition that allows such methods to
interact and to share information so that their power of inference
could be amplified. Concretely, we developed the notion of state that
captures computational state within a composition context and defined
partial solvers as certain state transformers. Two composition
operators for partial solvers were developed and shown to preserve
polynomial-time computability: a sequential iteration of a list of
partial solvers that tracks progress, and a lift operator testing
soundness of edge removals by exploring
consequences of edge commitments for a partial solver.

We instantiated these composition operators with partial solvers from
the literature and applied them experimentally to study games that
such composed partial solvers cannot simplify. These games, seen as data,
led to the incremental design of new partial solvers, even to
a new method that merges nodes known to have the same but
unknown winner. We proved the soundness of these new solvers.
Our focus was on computing winning regions, not winning strategies.
Would could compute finite-memory winning strategies in principle; it would be interesting to learn whether this could be done here for memoryless winning strategies as well.

We unit tested the implementation of our approach to validate experimental
results. The latter demonstrated the effectiveness of such a sequence
of refined partial solvers: after only a few refinement steps we
arrived at a partial solver that not only solved all structured games from the
state-of-the-art benchmark suite for parity games, but whose lifted
version also solved all random games generated within a
month of calendar time. We think this is compelling evidence that
there are very effective polynomial-time partial solvers for parity games.

The strength of this work is that is yields effective partial solvers
that are guaranteed to run in polynomial time. But this is also its
weakness in that we do not, at present, have a good understanding of
what types of parity games are solved completely for certain partial
solvers. More powerful versions of Theorem~\ref{theorem:oneplayer}, which extend to classes
of 2-player games, would be a first step in addressing that
weakness.

\section*{Open Access to Research Data}
Our source code is openly accessible at 

\begin{center}
\verb+bitbucket.org/Ah-Fat/gandalf_source+
\end{center}

\noindent Structured benchmarks we used were not our own and are accessible through the references provided in this paper. We chose not to store the random games we generated. This is justified by the fact that the random generators are publicly available and so these experiments can be repeated in principle on freshly generated random input, where the expectation is that results will be similar in quality.

\bigskip\noindent
{\bf Acknowledgements:} We thank Nir Piterman very much for his comments on
this work.

\nocite{*}
\bibliographystyle{eptcs}

\end{document}